\pdfoutput=1
\documentclass[twocolumn, switch]{article} % Method A for two-column formatting

\usepackage{preprint}

\usepackage{amsmath, amsthm, amssymb, amsfonts}
\usepackage{multirow}
\usepackage{caption}
\usepackage{subcaption}
\usepackage[numbers,square]{natbib}
\usepackage{tabularx}

\usepackage[utf8]{inputenc}	% allow utf-8 input
\usepackage[T1]{fontenc}	% use 8-bit T1 fonts
\usepackage{xcolor}		% colors for hyperlinks
\usepackage[colorlinks = true,
            linkcolor = purple,
            urlcolor  = blue,
            citecolor = cyan,
            anchorcolor = black]{hyperref}	% Color links to references, figures, etc.
\usepackage{booktabs} 		% professional-quality tables
\usepackage{nicefrac}		% compact symbols for 1/2, etc.
\usepackage{microtype}		% microtypography
\usepackage{lineno}		% Line numbers
\usepackage{float}			% Allows for figures within multicol

\usepackage{lipsum}		%  Filler text

\usepackage{newfloat}
\DeclareFloatingEnvironment[name={Supplementary Figure}]{suppfigure}
\usepackage{sidecap}
\sidecaptionvpos{figure}{c}

\usepackage{titlesec}
\titlespacing\section{0pt}{12pt plus 3pt minus 3pt}{1pt plus 1pt minus 1pt}
\titlespacing\subsection{0pt}{10pt plus 3pt minus 3pt}{1pt plus 1pt minus 1pt}
\titlespacing\subsubsection{0pt}{8pt plus 3pt minus 3pt}{1pt plus 1pt minus 1pt}

\usepackage{tikz,xcolor,hyperref}

\definecolor{lime}{HTML}{A6CE39}
\DeclareRobustCommand{\orcidicon}{
	\begin{tikzpicture}
	\draw[lime, fill=lime] (0,0)
	circle [radius=0.16]
	node[white] {{\fontfamily{qag}\selectfont \tiny ID}};
	\draw[white, fill=white] (-0.0625,0.095)
	circle [radius=0.007];
	\end{tikzpicture}
	\hspace{-2mm}
}
\foreach \x in {A, ..., Z}{\expandafter\xdef\csname orcid\x\endcsname{\noexpand\href{https://orcid.org/\csname orcidauthor\x\endcsname}
			{\noexpand\orcidicon}}
}
\title{Phishing Codebook: A Structured Framework for the Characterization of Phishing Emails}

\author{
  Tarini Saka \\
  University of Edinburgh \\
  Edinburgh, United Kingdom \\
  \texttt{tarini.saka@ed.ac.uk} \\
   \And
  Rachiyta Jain \\
  \texttt{rachiyta.j@gmail.com} \\
   \And
  Kami Vaniea \\
  University of Waterloo \\
  Waterloo, Canada \\
  \texttt{kami.vaniea@uwaterloo.ca} \\
   \And
  Nadin K\"okciyan \\
  University of Edinburgh \\
  Edinburgh, United Kingdom \\
  \texttt{nadin.kokciyan@ed.ac.uk} \\
}

\begin{document}

\twocolumn[ % Method A for two-column formatting
  \begin{@twocolumnfalse} % Method A for two-column formatting

\maketitle

\begin{abstract}
Phishing is one of the most prevalent and expensive types of cybercrime faced by organizations and individuals worldwide. Most prior research has focused on various technical features and traditional representations of text to characterize phishing emails. There is a significant knowledge gap about the qualitative traits embedded in them, which could be useful in a range of phishing mitigation tasks. In this paper, we dissect the structure of phishing emails to gain a better understanding of the factors that influence human decision-making when assessing suspicious emails and identify a novel set of descriptive features. For this, we employ an iterative qualitative coding approach to identify features that are descriptive of the emails. We developed the ``Phishing Codebook'', a structured framework to systematically extract key information from phishing emails, and we apply this codebook to a publicly available dataset of 503 phishing emails collected between 2015 and 2021. We present key observations and challenges related to phishing attacks delivered indirectly through legitimate services, the challenge of recurring and long-lasting scams, and the variations within campaigns used by attackers to bypass rule-based filters. Furthermore, we provide two use cases to show how the Phishing Codebook is useful in identifying similar phishing emails and in creating well-tailored responses to end-users. We share the Phishing Codebook and the annotated benchmark dataset to help researchers have a better understanding of phishing emails.
\end{abstract}

\end{@twocolumnfalse} % Method A for two-column formatting
]

% keywords can be removed
\keywords{Phishing\and Phishing Campaigns\and User Guidance\and Email Security\and Qualitative Study\and Codebook}

\section{Introduction}

Phishing is a type of cyberattack in which criminals impersonate legitimate organizations and send messages via email or other means to get the user to perform an action against their best interest, such as providing sensitive login information. Phishing is one of the most common and disruptive cyberattacks faced by organizations~\cite{apwg2023, cybersecurityukgov,verizon2021} and is one of the four major entry points into an organization~\cite{verizon2022}. According to the IBM Security X-Force Threat Intelligence Index 2023, phishing is the leading infection vector for security breaches, identified in 41\% of the incidents~\cite{ibmSecurityXForce}. The credentials obtained through phishing are often used to launch more damaging attacks such as ransomware or privilege escalation. Phishing attacks often result in significant financial and reputation damage to organizations and individuals alike. Given the low operational cost and high potential rewards, phishing attacks are becoming increasingly popular, causing a massive burden on organizations and security researchers. 
%Hence, organizations have dedicated extensive resources and research to combat the substantial danger posed by phishing and mitigating the damage inflicted.

Owing to advances in machine learning~(ML) and artificial intelligence~(AI), technological defenses are becoming more effective at automatically identifying and blocking malicious content~\cite{fette2007, benavides2020, khonji2013}. However, attackers are constantly creating new, highly sophisticated, and personalized attacks, which makes relying solely on detection techniques insufficient. As a result, users continue to encounter phishing emails in their mailboxes. These emails manage to bypass email filters and security scans, making them even more challenging to mitigate. Phishing emails often use various tricks to disguise their true source, such as vaguely stating that the email is from ``IT Services'' with the implication that the email is from the user's organization. Additionally, an email can be visually formatted to imply an email's source without ever naming the impersonated organization. These tactics can deceive most humans and also obfuscate the implied email source from computers. However, end users with security expertise can often identify and report these emails by analyzing a combination of technical (URLs and Sender information) and descriptive features (Threats and Cues). This is due to the difference in how humans and computers perceive and process emails. It is currently unclear how organizations could leverage the strengths of both humans and computers to combat phishing attacks.

We focus on combining ML algorithms with human expertise to create a hybrid defense system. While technical aspects of phishing emails have been thoroughly studied and automated, descriptive aspects are still under-explored and hardly automated. Integrating qualitative user-perspective features into phishing email detection models could also enhance the explainability of ML algorithms by aligning the model's decision-making process with human-recognizable patterns. To develop an effective hybrid defense system, we need a structured representation of qualitative features in phishing emails that influence human decision-making when assessing potentially malicious emails. For this, we conduct an iterative qualitative analysis of a publicly available set of phishing emails to introduce the \textit{Phishing Codebook}, a set of high-level descriptive features in phishing emails. 
%\note{Integrating qualitative user-perspective features into phishing email detection models enhances the explainability of ML algorithms by aligning the model's decision-making process with human-recognizable patterns. This alignment allows users and developers to understand and trust the model's outputs, as the reasoning behind classifications is more transparent and relatable to human intuition.}
In addition, we explore how the codebook can be utilized in two phishing mitigation tasks: (A) identifying phishing campaigns, and (B) providing guidance/education to users. The former often involves grouping similar emails based on the underlying fraud, making them easier and faster to process by IT staff~\cite{saka2022,althobaiti2021}, or identifying common origins~\cite{seifollahi2017, alazab2013}. The latter ensures that any malicious emails in mailboxes are reported, preventing the attack from escalating into a data breach within the organization.

%To address~(A), human experts can provide valuable insights into the context of emails, such as language and sender reputation, which can in turn be used to enhance the performance of ML algorithms. To address~(B), expertise gained from security researchers can be used to assist non-expert users. For instance, an assistive AI tool could provide contextual advice to users when they report a suspicious email, such advice needs to be quick and tailored to the specific email to be useful.

In this paper, we apply our Phishing Codebook to a dataset of 503 publicly accessible emails collected over a period of seven years and release this benchmark dataset for future research~\footnote{We will share the dataset link upon publication.}. We analyze this dataset to provide an in-depth analysis of emerging patterns and challenges observed in phishing emails individually and also in a group (i.e. campaigns). To address the two mitigation tasks mentioned above, we illustrate how the qualitative features can be beneficial to group phishing emails, and discuss how they can be used to provide better guidance to end-users. In summary, our paper addresses the following research questions: 
%The insights gained from this study can help researchers create stronger defenses against phishing attacks by combining human-like decision-making with machine capabilities. 

\begin{enumerate}
    \item[\textbf{RQ1}:] What qualitative attributes can we observe in phishing email text that replicate human perception?

    \item[\textbf{RQ2}:] To what extent do these qualitative attributes help in identifying phishing email campaigns?
    
    \item[\textbf{RQ3}:] What kind of observations or challenges emerge from a study of phishing email content?
\end{enumerate}

The primary outcome of this study is the Phishing Codebook, as described in Section~\ref{sec:codebook}, along with the results derived from the coded dataset. Section~\ref{sec:obvs} outlines the various observations and challenges posed by phishing emails, highlighting the improper use of legitimate services to trick users, the persistence of recurring scams that are difficult to detect over time, and the challenges arising from the variability within campaign emails. Furthermore, we discuss the application of the Codebook in two distinct use cases: campaign detection and user guidance (Section~\ref{sec:applications}). Section~\ref{sec:disscussion} provides a summary of our findings and limitations, and we conclude with Section~\ref{sec:conc} to discuss our future directions.

\section{Background}
\label{sec:bg}

The main focus in existing phishing literature is the automatic detection of phishing attacks. Khonji~\textit{et al.}~\cite{khonji2013} analyze existing work on the detection of phishing attacks and provide a high-level overview of various categories of phishing mitigation techniques: blacklists, rule-based heuristics, visual similarity, and machine-learning-based classifiers. Based on their analysis, ML-based detection techniques achieve high classification accuracy for analyzing similar data parts to those of rule-based heuristic techniques~\cite{fette2007, smadi2018, muralidharan2023}. Fette~\textit{et al.}~\cite{fette2007} develop a support vector machine-based algorithm, called \textit{PILFER}, to classify an instance as phishing or not using a set of ten binary and continuous numeric features. Smadi~\textit{et al.}~\cite{smadi2018} provide a reinforcement-learning-based solution to detect phishing emails. Muralidharan and Nissim~\cite{muralidharan2023} present a fully automated malicious email detection framework that uses deep learning methods to analyze all segments of the email (body, header, and attachments). Although these algorithms show promising results, one cannot expect that a system could provide $100$\% accuracy. Consequently, users still receive many phishing emails in their inboxes. The presence of an active and adaptive adversary is the primary reason that phishing attacks remain prevalent despite efforts to stop them. Hence, dedicated efforts are required to deal with undetected phishing emails that reach the inboxes of end-users, block mass attacks, and provide timely response/advice to users.

\subsection{Grouping Similar Phishing Emails}
One of the most effective phishing mitigation techniques is identifying phishing campaigns and taking down the phishing resources~\cite{khonji2013}. This is often achieved by reporting attacks to Internet Service Providers or IT Security teams. The identification of phishing email campaigns remains an under-explored research area because of the lack of consistency in the characterization of campaigns. Identifying phishing campaigns is very useful because an organization can promptly delete and blacklist all instances immediately after detecting a single element by accurately identifying all emails linked to a particular phishing campaign. Such techniques have been previously utilized to identify spam campaigns~\cite{xie2008, sheikhalishahi2020, saka2022} and phishing website campaigns~\cite{layton2010, layton2012}, and have proven to be effective tools in combating cybercrime. 

There are several use-cases for grouping a set of similar phishing emails. For example, Xie \textit{et al.}~\cite{xie2008} identify spam email campaigns to \textit{identify and take down spam botnets}. Similarly, Calais Guerra~\textit{et al.}~\cite{guerra2008} present Spam Miner, an online system designed for real-time monitoring and characterization of spam traffic over the Internet. Their system was used by the Brazilian Network Information Center and it mined more than 350 million spam messages, detecting meaningful clusters and patterns, and helping the organization to better understand the spam problem. In~\cite{saka2022} and~\cite{althobaiti2023}, the authors show that grouping similar emails could assist computing support teams in \textit{managing the high volume of phishing reports} that follow a large attack against an organization. Another common application of email grouping is \textit{authorship attribution}~\cite{seifollahi2017}, \cite{alazab2013}. These works use authorship attribution techniques to profile different attack behaviors that match offenders or a crime group and provide a consistent method to track groups over time, even if some characteristics of the emails change. In each of the above studies, researchers had different motivations but the common goal was to find meaningful clusters or groups of phishing emails. However, these groups might not necessarily be actual campaigns. We will be using the following definitions throughout the paper.

\begin{enumerate}
    \item \textbf{Phishing email scam:} Broadly, this refers to the hoax that the phishing email contains. Examples include package delivery and bank account scams.
    \item \textbf{Phishing email campaign:} A large number of emails sent by a common source as part of a single attack. Member emails share common characteristics like the underlying scam, the malicious element (URL or attachment), organization named and so on. However, the range of variation and commonalities among emails differ from campaign to campaign.
    \item \textbf{Phishing email cluster:} A group of phishing emails that exhibit notable similarities or patterns, regardless of their source or origin. For instance, emails based on the same template or kit need not be from the same attacker.
\end{enumerate}

\subsection{ML-based Representation of Phishing Emails}
One of the key steps in automated phishing techniques is defining a structured representation of phishing emails. The appropriate feature set would require consideration of all parts of the emails such as the \textit{email header}, \textit{email body}, \textit{attachments}, and \textit{URLs} included in the email body~\cite{fette2007, ma2009}. Of these, the body text is particularly important for campaign detection and user guidance as it provides the context for the email. Several studies have demonstrated the efficiency of using email body text to enhance the performance of phishing classifiers~\cite{toolan2010, saka2022, alhogail2021, song2010, abu2007}. However, the highly unstructured nature of email body text and the significant variation in language, grammar, and layout present significant challenges for machine interpretation. Traditional ways to represent the email body text primarily used syntactic and linguistic features of the text~\cite{dinh2015, alazab2013, park2015}, but these fail to fully encapsulate the context of an email. More recently, researchers have turned to advanced natural language processing (NLP) and large language models to better represent the contextual nature of email body text~\cite{alhogail2021, bountakas2021, saka2022}. These methods offer a more nuanced understanding of the text, enabling a more effective classification of phishing emails or identification of campaigns, which may change significantly over time. However, the feature set required to identify a campaign or design response should be descriptive of the context. There is potential for using qualitative features that we could learn from humans that are descriptive of the email rather than quantitative ones. 

\subsection{Human-centric Representation of Phishing Emails}
Humans play a crucial role in the phishing ecosystem and act as the last line of defense against phishing attacks. Depending on the level of technical expertise, some users can spot even the most sophisticated and well-crafted attacks. This can be attributed to the intricate cognitive processes that humans go through when evaluating an email. Wash~\textit{et al.}~\cite{wash2021} present evidence that humans read and judge emails in a very different way, using different knowledge and capabilities, than almost all of the technical filters. Through a survey of 297 non-expert participants, they show that the most common aspect noticed by people was that the email included a request for an action~(76\%). This corresponds to their finding that requests for actions (action links) were important triggers even for experts~\cite{wash2020}. The second most commonly noticed aspect of an email was `what the email was about', with 52\% of respondents noticing this. Although such aspects play a crucial role in the human cognitive process, they are not commonly used in phishing filters or campaign detectors, and when they are, they are often limited in scope by language issues~\cite{wash2021}. Although they are harder to extract, we argue that utilizing such aspects as features can significantly improve the performance of campaign detection algorithms. We attempt to define a robust set of human-facing features by creating the \textit{Phishing Codebook}.

The Phishing Codebook, while serving as an information extraction framework, resembles a taxonomy. Aleroud and Zhou~\cite{aleroud2017phishing} propose a phishing attack taxonomy comprising four dimensions: Communication Media, Target Environments, Attack Techniques, and Countermeasures. Although `Email Phishing' is included as a type of communication media in the taxonomy, it lacks detailed information and does not explore the structure of the email. Similarly, Gupta et al.~\cite{gupta2018defending} present a taxonomy focusing on two types of attacks: Social Engineering and Technical Subterfuge. Under social engineering, they define email spoofing. However, in their taxonomy, email phishing attacks are classified into two categories: spear-phishing and whaling. This is insufficient to represent the full landscape of email-based phishing attacks and the various tactics used.
Rastenis et al.~\cite{rastenis2020mail} propose a taxonomy specifically for email-based phishing attacks, consisting of six phases, and adapted it as a notation for an incident management system. The second stage is ‘E-mail content creation’, which contains four criteria: email presentation; text generation strategy; email text creator type; and email text personalization level. Although their taxonomy is comprehensive, it focuses on the overall journey of a phishing email, with limited qualitative and descriptive features identified. 
Singh\textit{ et al.}~\cite{Singh2021Hard} identified features that make phishing emails hard to detect: sender mismatch, request for credentials, urgency, offers, suspicious subjects, and link mismatches. Each feature had a binary result (present or not). Their study highlights the need for targeted training on these deceptive tactics. %Our study adds an additional dimension to this research, focusing on the context of email. We aim to create a structured format for the email context and provide an efficient representation.
Another important study is the PhishScale~\cite{steves2020phishscale}, which was developed by NIST to measure the click rate of phishing simulation based on the complexity of phishing emails. Although there are similarities in the features used, reflecting common literature and experience, our Phishing Codebook focuses on modeling the context of phishing emails for practical use. While PhishScale provides a wide range of features, it was designed for manual use as many of the features cannot currently be automated based on current research. The Phishing Codebook prioritizes features that replicate human thought processes and could potentially be automated.

\subsection{Automatic Response to Users}
It is important to provide timely and suitable guidance to users to keep them safe from phishing attacks, which often use urgent or threatening language to trick users into taking dangerous actions. NCSC recommends that employers train their employees to identify and report suspected phishing emails as part of a multi-layered approach to security~\cite{ncsc2021}. Crowd-sourced phishing reporting provides fast detection, small operational overhead, and sustained employee reporting activity~\cite{lain2022}. However, employee training is often provided up-front~\cite{redmiles2016learned} or after an employee falls for mock phishing~\cite{kumaraguru2010teaching} and is not always effective at the moment. Zhuo~\textit{et al.}~\cite{zhuo2023sok} propose a three-stage Phishing Susceptibility Model (PSM), which covers three temporal stages that explain human vulnerability to phishing attacks. One of their key findings is the research gap around in-the-moment assistant tool effectiveness. To address this issue, Jenkins~\textit{et al.}~\cite{jenkins2022phished} propose a design for a novel phishing-advice tool, PhishEd, to provide quick and accurate support to those who report phishing attempts. They propose the use of various contextual features, URLs, and email headers to craft appropriate responses and support users in making informed decisions. By understanding the qualitative features that experts use, we can provide better responses to end-users and support them in making informed decisions.
\section{Qualitative Analysis of Phishing Emails}
\label{sec:method}

%Few studies attempt to combine the perspectives of AI and Human Factors when considering phishing emails. Many studies look at the computer-friendly features of phishing emails~\cite{althobaiti2019review, khonji2013, toolan2010}. Similarly, other work has considered how people interpret the content of phishing emails~\cite{pilavakis2023,butavicius2022}. However, there have been few works that aim to cross this divide by creating a structure that allows AI work to consider how people interpret phishing emails.  

Qualitative research is a method of inquiry that involves collecting and analyzing non-numerical data, such as text, audio, or video, to gain insights into concepts, opinions, or experiences~\cite{saldana2016}. It has been used in various fields, including medicine~\cite{sibeoni2020}, social sciences~\cite{brinkmann2014}, and usable security~\cite{pilavakis2023}, and has led to valuable contributions to our understanding of many real-world problems. In this study, we employ the approach of \emph{qualitative coding} to minimize the subjectivity in human judgment and create a structured representation of phishing emails that is likely to be reproducible by other researchers. Through an iterative coding process, we analyzed a dataset of phishing emails and identified important attributes in email texts that humans would naturally observe. In this section, we outline the methodology used in this study. First, we establish a systematic framework for extracting critical information from phishing emails, enabling us to create a structured consistent representation of their content. Second, we build an annotated dataset that captures the qualitative characteristics of phishing emails, which can provide valuable insights into the strategies employed by malicious actors and contribute to the development of effective countermeasures to combat phishing.

\subsection{Dataset}
We use the \textit{Nazario Phishing Dataset}~\cite{nazario}, a publicly available collection of hand-screened phishing messages collected by Jose Nazario from his personal inbox. It is the most well-known phishing email dataset available and has been previously used in multiple works for phishing detection and classification~\cite{bountakas2021, saka2022, zeng2020}. While not necessarily representative of what other users might encounter, we consider this a suitable dataset for studying user-oriented aspects for two reasons. First, it is publicly available so we can release the annotated benchmark dataset publicly. Datasets from organizations or security providers are often restricted by Non-Disclosure Agreements (NDAs), preventing public dissemination. Second, this dataset spans over 7 years (2015 to 2021), allowing us to accommodate for temporal shifts in phishing attacks and identify reliable features applicable to a wide range of emails. All emails collected in a single year were published in a single collection of email messages in plain text (i.e., mbox) and were downloaded from the original site.

\textbf{Pre-processing.} The email files were analyzed using the \texttt{mailbox} module in Python. We extracted all relevant features and created a spreadsheet for the entire dataset. A unique ID was assigned to each email, consisting of the year and the index of the email in the initial dataset (e.g., $2015\_001$). Any emails with empty content ($n=201$) and non-English content ($n=114$) were removed. The final dataset consisted of 1,688 processed emails.

\subsection{Coding Methodology}
The qualitative coding process was conducted in two parts. In the first, \textit{exploratory phase}, we iteratively developed the phishing codebook, which was then used for the \textit{main coding phase}.

\subsubsection{Unit of analysis}
Codes were applied to the user-visible portion of the email, rather than just the text passages. Coders ignored the content of headers beyond what a normal email client would show. This means that information like DKIM signatures were not considered, but \emph{from} and \emph{subject} were included. Email attachments were not opened for safety reasons, and to streamline the coding process. Due to the age of the emails, URLs were not clicked as most no longer show the original content, and clicking on them would be unsafe for the coders and slow down the coding process. We also want to ensure that the code is efficient and safe for other coders, especially for researchers who may lack security knowledge. Note that information like the existence of attachments and the domains of links can be automatically extracted, and hence were not considered in the manual codebook.

\subsubsection{Exploratory phase}
\label{sec:phase1}

Three authors engaged in an iterative coding process comprising five rounds. The primary objective of these rounds was to gain a comprehensive understanding of the dataset and to refine the coding scheme. The authors included a PhD student in AI, a master's student in Law, and a professor with expertise in human-factors of phishing. The first author identified a 1-2 month period with the maximum email frequency in each year and randomly sampled emails from this period. This approach provided a representative sample of phishing emails across different periods and formed our preliminary dataset. The first round was unstructured, similar to open coding or memoing, with the authors familiarising themselves with the dataset by reading through emails, noting interesting features, and considering suitable coding methods to identify aspects of phishing emails likely to be readily salient to general users. We reviewed 40 emails from the preliminary dataset and then discussed the observations.

Based on the first round of discussion, we selected a hierarchical coding method and identified key elements in emails that were descriptive of the context, consistent with time, and clear for coding. This included the claimed source, salutation, threatening or urgent cues, actions, and claimed purpose of the email. 
The claimed email source is an important factor when judging phishing but many emails leave it implied (e.g.\ ``your organization'' or ``IT Services'') making it hard to code correctly. So we break it into named organizations, when present, and sectors which were often easier to discern. Similarly, the requested action is quite important, particularly when giving users advice, so we broke it into high-level actions (e.g. ``click a link'') and then in-vivo coded the specific action goal (e.g.\ ``update account''). These mixes of clear classifications with open-ended in-vivo coding approaches are meant to make the codes easier for automated learning while also keeping the human-perspective.

The codebook was then refined by applying it to 40 new emails, followed by a discussion on any points of disagreement or confusion, leading to further refinement. This iteration was repeated four times until the codebook stabilized, and the inter-rater reliability showed good agreement. After stabilization a fifth round was completed, resulting in an average inter-rater reliability of $0.75$ over all the codes. The lead author then manually compared the annotations and identified points of disagreement, which were then thoroughly discussed. For instance, in one email the action was not explicitly stated but implied that the recipient should call a number. One coder coded it as `call' and another coded it as `none'.  After discussing this case, we decided to focus the code on explicitly stated actions as they are less subjective. We then updated the codebook by either adding more details to the instructions or including an example. We followed a similar process for all points of disagreement.

\subsubsection{Main coding phase}
For the main coding, we created a dataset of 503 phishing emails, sampled using a frequency-based algorithm. To create a representative and diverse subset, we sampled approximately the same number of phishing emails from each year over the seven-year span. For each year, we identified the 3-4 month period with the maximum email frequency and sampled 72 emails from it. Our sampling choice was based on multiple factors. (1) We wanted to sample emails from a long time period so as to identify features that are reliable with the temporal evolution of phishing, and (2) We wanted to capture some phishing campaigns so we could additionally study them from a qualitative perspective. High-frequency periods (3-6 months in our study) increased the chances of capturing campaigns, as previous studies observed~\cite{drury2022, alazab2013}. We prioritized gathering sufficient data for reliable and reproducible feature identification. Due to a smaller number of emails available from 2020 and 2021, a larger time frame was used. Table~\ref{tab:dataset2} provides details on which months were sampled for each year, the number of emails in those months, and how many were sampled.

\begin{table}[h!]
 \caption{Summary of the coded dataset. The year, number of emails in the year, maximum frequency period, and number of emails sampled from the year. The identified period in 2020 comprised only 71 emails.}
 \label{tab:dataset2} 
 \centering
 \begin{tabular}{c c l c}
  \toprule
  \textbf{Year} & \textbf{Total} & \textbf{Max. freq. period} & \textbf{Sample Size}\\
  \midrule
   2015 &  307 &  Oct, Nov, Dec &  72 \\
  \midrule
   2016 &  494 &  Mar, Apr, May &  72 \\
  \midrule
   2017 &  325 &  Jan, Feb, Mar &  72 \\
  \midrule
   2018 &  288 &  Aug, Sep, Oct &  72 \\
  \midrule
   2019 &  243 &  Jan, Feb, Mar, Apr &  72 \\
  \midrule
   2020 &  158 &  Jan-Jun &  71 \\
  \midrule
   2021 &  101 &  Jan-Dec & 72  \\
  \bottomrule
 \end{tabular}

\end{table}

\subsubsection{Inter-Rater Reliability}
The first author coded the entire dataset and the second coder coded the last 50 emails, representing ($10\%$) of the dataset. This allowed us to assess inter-coder reliability and monitor any potential drift. We computed the agreement between the two coders using Krippendorff’s alpha~\cite{hayes2007} and Cohen’s Kappa~\cite{cohen1960} on each high-level code and calculated the overall average, $0.93$ Krippendorff’s alpha and $0.93$ Cohen's Kappa. These chance-adjusted indices are commonly employed to quantify the level of agreement among raters. Both coefficients have a range from $-1$ to $1$, where a value of 1 signifies perfect agreement.

These agreements are rather high for qualitative data, but make sense for this use case, as highly salient features were specifically selected to make a codebook that other researchers could use. Table~\ref{tab:reliability} provides Krippendorff’s alpha and Cohen's Kappa scores for each high-level code with sub-codes.

\begin{table}[htb]
 \caption{Inter-Rater Reliability Scores (Cohen's Kappa and Krippendorff's Alpha) for each of the high-level codes, where subcodes were pre-defined.}
    \label{tab:reliability}
\begin{center}
 \begin{tabular}{lcc}
 \toprule
 \textbf{High-Level Code} & \textbf{C. Kappa} & \textbf{K. Alpha}\\
 \midrule
  From – Company Name  &  0.96   &  0.96\\
 \midrule
  From – Sector      &  0.94   &  0.94\\
 \midrule
  Salutation           &  0.94   &  0.95\\
 \midrule
  Threatening Language &  0.96   &  0.96\\
 \midrule
  Urgency Cues         &  0.80   &  0.80\\
 \midrule
  Action - Generic     &  0.94   &  0.94\\
 %\midrule
  %Legitimate Source    &  1.00   &  1.00\\
 \bottomrule
\end{tabular}    
\end{center}
\end{table}

\section{Result - A Phishing Codebook}
\label{sec:codebook}

This section presents the main outcome of our qualitative study, the Phishing Codebook (as summarized in Table~\ref{tab:codebook}). We describe each code, emphasizing its role in email cognition and the coding criteria. This codebook addresses our first research question~(RQ1), as each of the eight high-level codes described corresponds to an attribute observed in phishing emails that replicates the human perspective. Additionally, we provide an overview of the statistical analysis conducted on the coded dataset. Table~\ref{tab:codebook} summarizes the eight high-level codes defined, providing a brief explanation of each code along with possible sub-codes. The codebook facilitates consistent and structured extraction of crucial information from phishing emails. Further details and instructions for coders are available in Appendix~\ref{app:app_codebook}. Note that normalizing email text and varying email lengths can be a challenge for machine learning~(ML) algorithms. To address this issue, we incorporate the ``Main Topic'' and ``Action-Specific'' codes in our codebook. These codes are designed to make the email text uniform, leading to enhanced clarity and conciseness. They can enhance the performance of various ML techniques that use textual features, such as topic models and text embeddings. We now explain each of the codes in detail while providing results from the dataset. Figure~\ref{tab:email_extraction} illustrates the categorization process of a email using our codebook.

\begin{table*}[htb]
 \caption{The Phishing Codebook. This table provides a summary of the final codebook developed, the eight high-level codes, a brief explanation of each, and predefined sub-codes.}
    \label{tab:codebook}
\begin{center}
    
 \begin{tabular}{lll}
 \toprule
 \textbf{High-Level Code} & \textbf{Explanation} & \textbf{Sub-Codes} \\
 \midrule
  From – Company Name &   Name of the organization being impersonated  & [in-vivo], organization, none\\
 \midrule
  \multirow{3}{*}{From – Sector}     &   \multirow{3}{*}{Type of sector the email claims to be from}  & financial, email, document share, \\&& logistics, shopping,                   service provider, \\&& security, government, unknown\\
 \midrule
  Salutation          &   Type of salutation used to address the recipient  & name, email, generic, none\\
 \midrule
 Threatening language &   Presence of threatening language, tone  & threat, none\\
 \midrule
 Urgency Cues         &   Presence of time pressure or urgency cues  & urgent, none\\
 \midrule
 Action – Generic     &   The action being prompted in the email  & click, download, reply/email, call, other, none\\
 %\midrule
 %Legitimate Source    &   The email is from a legitimate source   & legit, none\\
 \midrule
 Action – Specific    &   The reason provided to perform an action  & in vivo coding\\
 \midrule
  Main Topic          &   Main purpose of the email  & in vivo coding\\
 \bottomrule
\end{tabular}
\end{center}
\end{table*}

\begin{itemize}

\item \textbf{From – Company Name.} Phishers will often try to impersonate a real company using approaches that are easy for humans to interpret but potentially hard for computers. We code the organization name the email claims to be from (e.g., `Paypal') based on the sender address, subject line, or email body text. Only organization names are coded, and references to named persons are ignored. If the email claims to be from the user's organization without naming it, such as claiming to be from HR but not stating the organization name, we code it as simply `organization.’ If the email makes no claim about who it is from, or the coder cannot infer a specific organization name, then it is coded `none.’ 

\textit{Results from the dataset:} 74 distinct organization names were found in the email dataset. The majority of emails (27.43\%) had no definitive organization name. The top four organization names were; `monkey' (18.09\%), `USAA' (8.75\%), `Paypal' (6.16\%), and `WeTransfer' (2.98\%).  Followed by `organization' (2.78\%) where no specific organization was named.  The email provider that Nazario used to collect the phishing emails is `monkey.org' which is why it is so common,  phishers often name the email domain in messages.

\item \textbf{From – Sector.}
Even if the organization is unnamed, a sector is often implied with the hope that the human will fill in the details. For example, claiming that a package needs a fee for delivery without naming the delivery company. We chose nine sub-codes based on our observation: `financial', `email', `document share', `logistics', `shopping', `service provider', `security', `government', and `unknown'. If a company is involved in several different types of sectors, we code for the sector that the email content refers to. If the coder could not infer a sector, then we code `unknown'. Later applications of this codebook may need to add additional sectors depending on their corpus.

\textit{Results from the dataset:} The most common sector was `email' (45.13\%) which claim to be from the user's email provider. The second most common was `financial' (30.02\%), which included emails regarding banks and credit cards. 30 emails~(5.97\%) were coded as `unknown'.
 
Compromising email inboxes is highly valuable~\cite{krebsEmailValue} and often the first step to a larger compromise. In large organizations, an email password is often also used for other organization access, such as with Microsoft 365 services. Compromising an email account allows hackers to conduct internal attacks, such as distributing malware, launching ransomware attacks, and so on. Financial services have also historically topped the list of most impersonated industries in phishing~\cite{vade2022, apwg2023}.

\item \textbf{Salutation.} The type of salutation used to address the user. Some organizations like banks claim to always address the user by name, so phishers sometimes personalize emails by adding users' names or emails to make them seem legitimate~\cite{parsons2016}. We code if and how the email addresses the user at the top of the email body text, such as `Dear Jose' or `Attention user.’ 
%During the coding process, we observed emails without salutations but mentioned the recipient’s email address later in the content. For the sake of this study, we only consider names used at the top of the email in a traditional salutation format. 
There are four sub-codes in this high-level code: `name', `email', `generic', and `none'.

\textit{Results from the dataset: } 34.00\% of emails had no salutation, 31.41\% had a generic salutation (e.g.\ `dear user'), 21.67\% used the user's email address, and 12.92\% used the user's name.

\item \textbf{Threatening Language.} Phishers use threatening language to scare users and convince them to take action~\cite{williams2018, vishwanath2011}. Here, we code whether the email body text or subject line contains any explicit threatening language or tone. This includes talking about any negative consequences or losses of some kind such as \textit{`your files will be deleted'}. Indirect or implied threats were excluded. There are two sub-codes: `threat' and `none'.

\textit{Results from the dataset:} 55.67\% of emails had no threatening language and 44.33\% contained threatening language.

\item \textbf{Urgency Cues.} Phishers use urgency language or cues to create a sense of urgency or add time pressure to encourage, or even demand, immediate action in a bid to fluster the user, a tactic that is known to be effective~\cite{vishwanath2011, butavicius2022}. We code if the email message or subject line contains urgency cues such as time pressures, deadlines, and expiry dates. There are two sub-codes: `urgent' and `none'.

\textit{Results from the dataset:} 64.81\% of the phishing emails had no urgency cues and 35.19\% of the phishing emails contained urgency cues. An interesting observation made during the coding was the co-occurrence of threatening language and urgency cues, such as \textit{`Activation expires after 24 hours...and your domain monkey.org will be blocked'}. 142 emails contained both of these codes. Such instances formed approximately 80\% of emails with urgency cues and 64\% with threats.

\item \textbf{Action – Generic.} Simply opening a phishing email is rarely dangerous, instead, they ask the user to perform an action that creates a dangerous situation such as giving away credentials or downloading malicious code. Here, we code what actions an email is trying to coax the user into performing. There are six sub-codes; `click', `download', `reply/email', `call', `other', and `none'.

\textit{Results from the dataset:} The most common action was `click', with 433 emails (86.08\%) containing a link the user is expected to click on. The next most common action was `download', with 67 emails (13.32\%) asking the user to download an attachment. Only one email was coded `none'. Upon further analysis, it was observed that this email was meant to scare the user about a fraudulent transfer and coax them into calling the number provided in the email, although the email did not explicitly ask the user to do so. Three emails had `other' actions. In all three cases, the user was asked to copy and paste a link or address. Seven emails had multiple actions.

\item \textbf{Action - Specific.} The \textit{Action - Specific} code looks at the specific action requested of the user and provides a view into what the user is being asked to accomplish through that action. Since the range of potential specific actions is quite large and the language used can be quite important, we chose to use in vivo coding where the researcher uses words from the email to summarize the requested action. For example `to verify password'. 

\textit{Results from the dataset:} The 5 most common action-phrases were `verify account' (15), `update account' (12), `proceed email update' (9), `get files' (9), and `sign account update personal information' (6) The most frequent terms found across all specific action phrases were `account' (142), `update' (90), `email' (59), `verify' (46), `view' (46), `upgrade' (39), `confirm' (36), `attached' (34), `information' (31), and `messages' (23).
    
\item \textbf{Main Topic.} This code is used to define the main purpose of the email. We again used in vivo coding as the topics are quite varied.  This code addresses a concern highlighted in previous research~\cite{saka2022, han2016} and our own observations in the data, that the varying sizes of emails (short vs. long) and excessive overlap in text and wording can confuse Machine Learning~(ML) algorithms.

\textit{Results from the dataset:} The 5 most common topic-phrases were `passwords expiring soon' (14), `pending incoming emails' (13), `received files via WeTransfer' (10), `monthly account estatements' (9), and `mailboxes almost full' (5). The most frequent terms found across all topics were ``account" (135), ``email" (56), ``incoming" (52),  ``pending" (45), ``new" (43), ``password" (31), ``emails" (29), ``received" (28), ``soon" (27), and ``security" (26). Overall, the results show that the most common phishing scams are regarding the users' email accounts, passwords, and incoming communications/files.

\end{itemize}

\begin{figure}[h]
    \centering
    \includegraphics[scale=0.35]{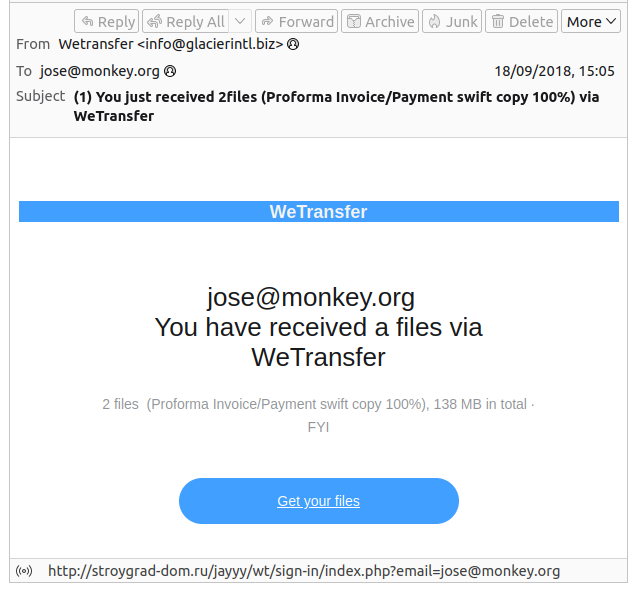}
    \caption{A example WeTransfer email.}
    \label{fig:your_image_label}
\end{figure}

\begin{table}[h]
    \centering
    \caption{Coded output of the email.}
    \label{tab:email_extraction}
    \begin{tabular}{l l}
        \toprule
        \textbf{Information Type} & \textbf{Email} \\
        \midrule
        From-Sector & document sharing \\
        Action-Generic & click \\
        Threat & none \\
        Urgency Cues & none \\
        From-Company & WeTransfer \\
        Main Topic & received a files \\
                   & via WeTransfer \\
        Action Specific & Get your files\\
        \bottomrule
    \end{tabular}
\end{table}

%\subsection{Summary}
%Why only these features? Discuss the representativeness and comprehensiveness of the features chosen.
%The proposed codebook contains eight high-level codes that capture the human-focused context of phishing emails. The codes ``From - Company Name" and ``From - Sector" offer insights into the impersonated organization, while ``Salutation" indicates the presence of personalization in the emails. ``Threatening Language" and ``Urgency Cues" highlight linguistic cues commonly used to influence the recipient's reaction and form an essential part of the scam. ``Action - Generic" and ``Action - Specific" describe the type of action being solicited from the recipient and the reason for the same. Finally, "Main Topic" summarizes the underlying scam, providing a concise description. The feature set in our codebook demonstrates a balanced approach to representativeness, incorporating diverse aspects of phishing emails while ensuring comprehensiveness by covering key dimensions relevant to the underlying scam of the phishing email. In the exploratory phase, we assessed other features like images, logos, and disclaimers. However, we encountered challenges in defining sub-codes for these elements, as categorizations such as `good', `mediocre', or `bad' proved to be highly subjective and posed difficulties in maintaining consistency among coders. 

\subsection{Application of the Codebook on an Independent Dataset}
\label{subsec:independent}

We applied the Phishing Codebook to a different dataset to determine whether it can be generalized to other organizations. The new dataset was collected from one department in a university in the United Kingdom. Students and staff were invited to ``donate'' phishing emails they received by forwarding them to a research inbox.  We selected 50 recent phishing emails forwarded by ten individuals from the period of November 2023 to May 2024 for analysis. The first author meticulously coded the entire set, and the third author subsequently reviewed all the codes, flagging any disagreements. These disagreements were then thoroughly discussed and resolved.

Overall, we found that the Phishing codebook worked well on the new dataset. The proposed high-level codes and sub-codes were easily adaptable, except for `From-Sector' where one new sub-code had to be added: `individual'. This refers to the emails sent from an individual offering a large sum of money as part of a charity, similar to the commonly observed Nigerian prince or 419 scam. Phishing emails collected from different sources will likely require some additions to the sub-codes of the proposed codebook, particularly for the `From-Sector'. The `individual' sector was the most frequently observed sector in this dataset. This is an interesting finding because emails from individuals typically contain only text and fewer detectable components, such as URLs. The most common action observed was `click,' consistent with findings from the Nazario set. A notable difference was that `Reply/Email' was the second most common prompted action, as opposed to 'Download' in the Nazario set. This could suggest that emails with attachments are better filtered out in the university, and the act of replying, which is commonly requested even in benign emails, is being used. The flexibility and adaptability of the proposed codebook make it generalizable across various contexts and organizations. Despite the need for minor adjustments, the primary framework remains intact and effective. This demonstrates that our codebook can efficiently categorize and analyze phishing attempts across different domains, maintaining its relevance and accuracy.

%-----------------------------------------------------------

\section{Observations and Challenges}
\label{sec:obvs}
This section presents some interesting observations about phishing attacks that emerged during the coding process, aimed at addressing our third research question~(RQ3). Additionally, we discuss some challenges often posed by phishing attacks and campaigns. Understanding such phishing trends can help design better detection algorithms, improve real-time monitoring, recognize phishing email templates matching known trends, and develop adaptive security measures that evolve to counter evolving phishing techniques.

\subsection{Improper Use of Legitimate Services}
\label{sec:legit}
Certain phishing attacks are delivered indirectly by routing them through authentic sources that send emails. For example, malicious documents can be shared through real file-sharing applications, commenting malicious links on real social media posts, and so on. These applications send notifications to users via email. Hence, the email appears to originate from a genuine and reputable organization, but it carries a concealed malicious element. 
The coders were asked to flag any such indirect phishing attacks for further analysis, resulting in nine emails identified. 
Subsequently, the lead researcher conducted a manual examination of the original \textit{.eml} files for these emails. All nine emails belonged to the document-sharing sector. Of these, two were found to be real emails from legitimate sources~(WeTransfer and DocuSign), but they likely lead the user to a malicious document.

Emails of this type, which misuse legitimate services, pose a complex security challenge. The emails themselves are real and devoid of any overtly malicious attributes or elements that security filters typically detect. They contain legitimate URL domains and sender addresses, IP addresses, and certifications. Consequently, we must depend on end-users to identify and report them, which requires educating users about such attacks specifically.

\subsection{Recurring and Long-lasting Scams}
%ORIGINAL EMLS: 2017_130, 2018_199
Despite using the same language and sharing similar wording and layout, similar scams reappear over multiple years. Existing detection algorithms have failed to filter them out. The \textit{WeTransfer} scam is a good example to discuss in detail to show why detecting scams is challenging. This scam impersonates WeTransfer, a widely used file-sharing service. Attackers send well-crafted emails to trick users into downloading harmful files instead of the expected shared content. WeTransfer scams could be categorized into two types: spoofed to appear to be from WeTransfer but links lead elsewhere, and legitimate emails from WeTransfer with an authentic link but presumably malicious content in the linked file. Section~\ref{sec:legit} shows that the latter is harder to automatically filter out due to the legitimate nature of the email.

In our dataset, we identified 15 WeTransfer scam emails occurring every year from 2017 to 2021. Only one email had an attachment to download; the rest contained a link for the user to click to ``get their files''. We analyzed each of these 15 emails in detail. Firstly, 6 of the 15 emails had the sender address domain \textit{wetransfer.com}, which is the real domain of WeTransfer. Domain spoofing is a common obfuscation tactic used to fool users~\cite{barracuda_spoofing}. The remaining emails had all different domains. With respect to URLs, one email had WeTransfer's real domain. This email with the real WeTransfer URL and the sender domain was likely a legitimate file-share email leading to a malicious document~(see Section~\ref{sec:legit}). Two emails shared a common top-level domain (\textit{web.app}, a web hosting platform) in their URLs, but had different specific URLs and sender addresses.

While we cannot definitively confirm whether these emails originated from the same source or belong to a single campaign, they are certainly similar enough for a detection algorithm to filter them. Such recurrent scams can be attributed to the growing availability of phishing kits. These off-the-shelf tools allow even non-technical criminals to spin up a phishing campaign and provide fully hosted, phishing as a service. Commonly employed defensive techniques, such as domain blacklisting and URL checkers, prove ineffective when attackers resort to tactics like spoofed domains, redirects, utilization of third-party web hosts, and the continual alteration of URLs. In light of these challenges, our study emphasizes the significance of integrating qualitative attributes, including the company name, the topic of communication, and the specific action requested, in conjunction with domain checks.

\subsection{Variation within Campaigns}
During our coding process, we observed several similarities and variations within emails that potentially belong to the same campaign (similar emails sent within a short time span). This finding aligns with previous research, that attackers often send similar emails in bulk~\cite{calais2008, althobaiti2023, dinh2015}, but often induce minor variations or obfuscations to bypass rule-based filters~\cite{calais2008, sheikhalishahi2020, alazab2013, althobaiti2023}. We observed a wide range of variations in technical, visual, and textual aspects.

\begin{figure*}[htb]
\begin{subfigure}{0.5\textwidth}
\includegraphics[width=\textwidth]{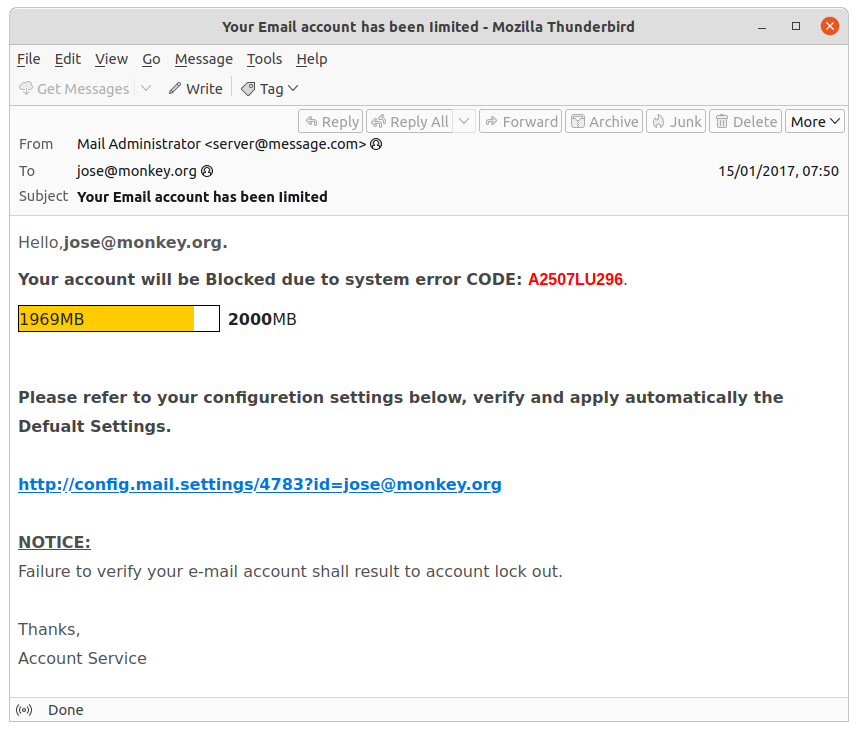} 
\caption{Email Sample 1}
\label{fig:subim1}
\end{subfigure}
\begin{subfigure}{0.5\textwidth}
\includegraphics[width=\textwidth]{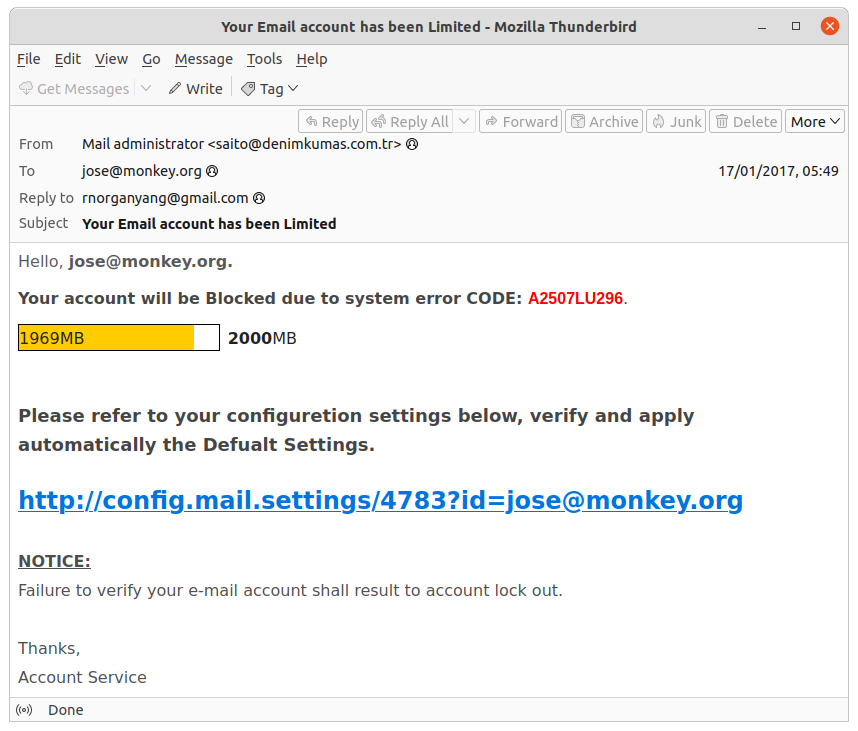}
\caption{Email Sample 2}
\label{fig:subim2}
\end{subfigure}
\caption{A comparison of two almost identical phishing emails sent to a user two days apart.}
\label{fig:same_ems}
\end{figure*}

For instance, two emails sent from a `Mail Administrator', sent two days apart, were compared and found to be visually identical, sharing similar subject lines, `error codes', and consistent grammatical errors (as depicted in Figure~\ref{fig:same_ems}). However, upon closer inspection, we noticed subtle variations in the subject line~(e.g., `Iimited' and `Limited') and different sender addresses. Furthermore, although the visible URLs in both emails appear identical, the embedded links reveal discrepancies even though they share a common top-level domain. Blocking the phishing URL and sender email are standard responses to phishing attacks. The high level of similarities and short time span between the emails indicate that they could belong to the same campaign. However, the technical disparities pose a formidable challenge for conventional grouping algorithms that use blacklists and rule-based filters.

 %ORIGINAL EMLS: 2017_018, 2017_020,  2017_021

\begin{figure*}[h]
\begin{subfigure}{0.32\textwidth}
\includegraphics[width=\textwidth]{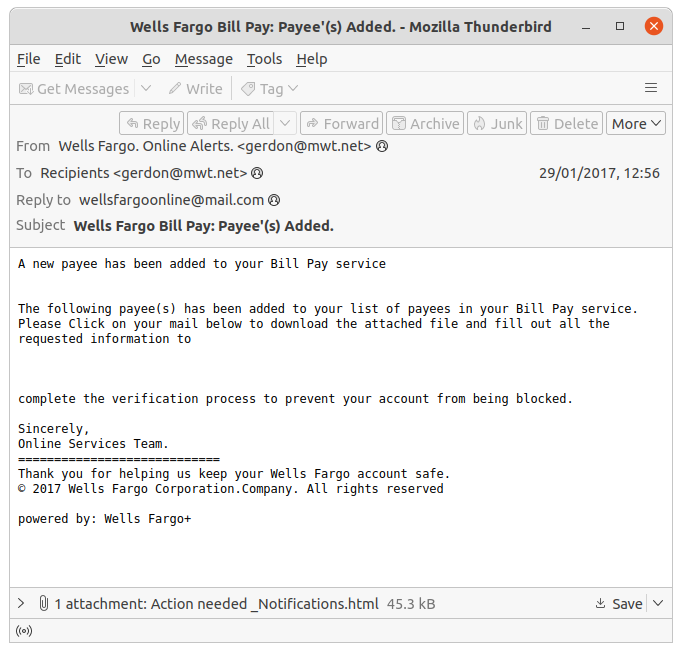} 
\caption{Wells Fargo Email}
\label{fig:sim_ems_1}
\end{subfigure}
\begin{subfigure}{0.32\textwidth}
\includegraphics[width=\textwidth]{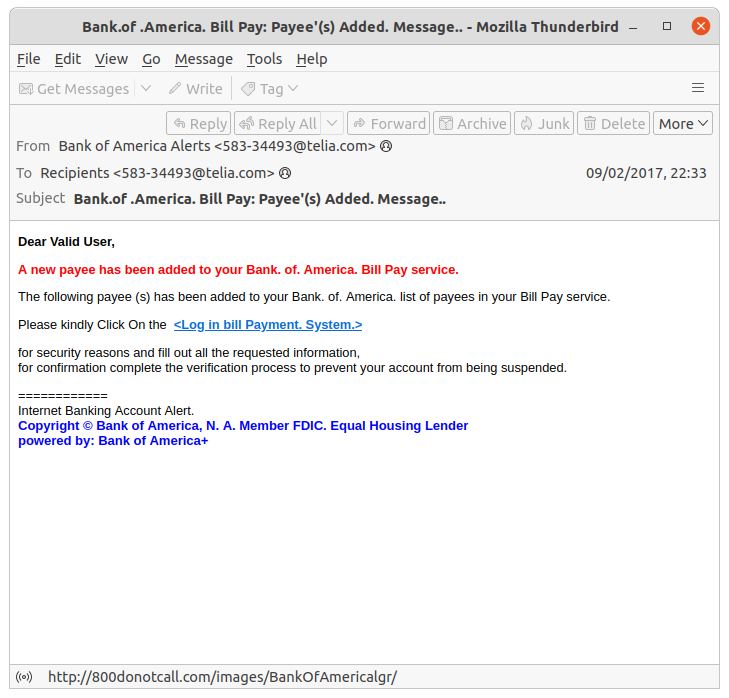}
\caption{Bank of America Email}
\label{fig:sim_ems_2}
\end{subfigure}
\begin{subfigure}{0.32\textwidth}
\centering
\includegraphics[width=\textwidth]{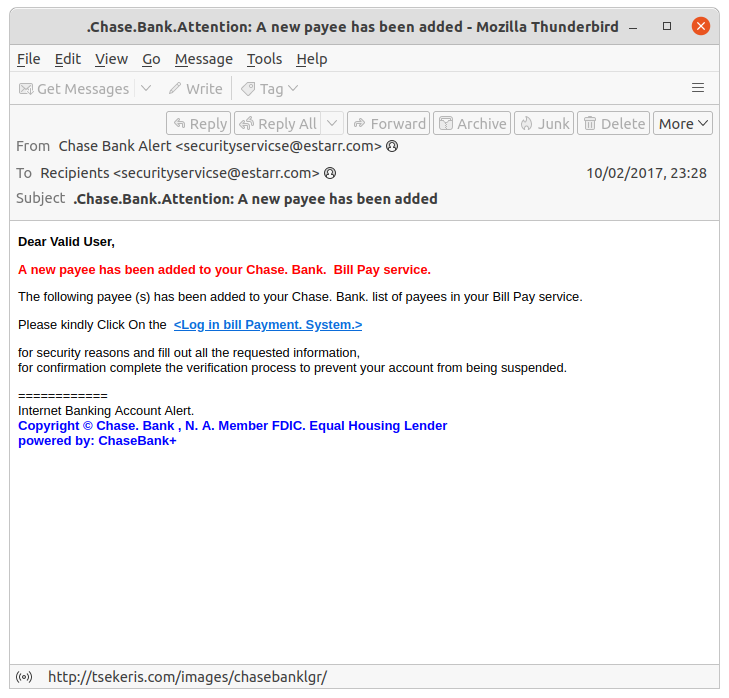}
\caption{Chase Bank Email}
\label{fig:sim_ems_3}
\end{subfigure}
\caption{A comparison of three similar phishing emails sent to a user impersonating three different banks.}
\label{fig:sim_ems}
\end{figure*}

In another instance, we examined three `New Payee' emails sent over two weeks, with two of them only two days apart. Two of these emails (Figure \ref{fig:sim_ems_2} and Figure \ref{fig:sim_ems_3}) were visually and textually identical, but also closely resemble the content of the first (Figure \ref{fig:sim_ems_1}). Despite claiming to be from different financial institutions (Wells Fargo, Chase Bank, and Bank of America), they share a common narrative about adding a new payee to the bill pay service. While each had distinct sender addresses and domains, one had an attachment, and the others had URLs with different domains. Although the subject lines were similar, they were not identical. Notably, the latter pair's close content match and timing suggest they may be part of the same campaign. However, differences in impersonated organizations, sender addresses, and URL domains pose challenges.

\section{Applications of the Phishing Codebook}
\label{sec:applications}
In this section, we will demonstrate the practical application of the proposed Phishing Codebook for two essential phishing mitigation tasks. These tasks are specifically designed to support two different user groups. Firstly, we will illustrate how the codebook's output can be used to identify similar phishing emails, likely from the same campaign. This can significantly enhance the IT staff's ability to manage reported phishing incidents efficiently. Secondly, we will discuss how the coded representation of the email can be used to create well-tailored responses to end-users and provide them with guidance on how to respond appropriately.

\subsection{Campaign identification}
\label{sec:app1}
To demonstrate the significance of using qualitative features to identify phishing campaigns~(RQ2), we conducted a multi-criteria classification experiment. The proposed classification approach is based on a small number of criteria, making it simple and efficient to implement, and thus useful for organizations that have limited resources for email security. Additionally, the classification approach can be easily adapted and extended to incorporate additional technical or qualitative features, to improve the accuracy. In Section 5.3, we show how phishers employ a range of obfuscation techniques to bypass email filters including minor variations in subject lines, sender addresses and domains, URLs, and organization names. Combining both qualitative and quantitative features from emails can considerably enhance the detection outcomes. For the purpose of this demonstration, we used five high-level codes and exact match criteria to group emails. In a larger experiment, we suggest researchers utilize the codes and matching criteria that they find most suitable for their specific domain and use-case.

\begin{table*}[t]
 \caption{Summary of the campaigns identified using our multi-criteria classification. Columns are campaign attributes (from-sector, company name, action-generic), main topic and action-specific, number of emails, year of the campaign, average Levenshtein distance of the subject line, visual match, number of unique From sender domains, and URL domains.}
    \label{tab:qual_camps}
\begin{center}
%\resizebox{\textwidth}{!}{  
 \begin{tabular}{lllclcccc}
 \toprule
 \textbf{Attributes} & \textbf{Main Topic and Action} & \textbf{No.} & \textbf{Years} & \textbf{Subject} & \textbf{Visual} & \textbf{From} & \textbf{URLs}\\
 \midrule
  email [none] & password expire soon & 6 & 2015 & 0.0 & same & 4 & 3 \\
  click & proceed email update &  & 2016 &  &  &  &  \\
  %click &  &  &  &  &  &  &  \\
 \midrule
  financial [nedbank] & monthly account estatement & 5 & 2018 & 19.7 & same & 2 & 1 (legit) \\
  download &  encrypted electronic statement &  & 2020 &  &  &  &  \\
  %download &  &  &  &  &  &  &  \\
 \midrule
  financial [usaa] & new security safeguard & 4 & 2015 & 20.3 & same & 2 & 4 \\
  click & update personal information &  &  &  &  &  &  \\
  %click &  &  &  &  &  &  &  \\
 \midrule
  document [wetransfer] & received file via wetransfer & 7 & 2018 & 30.3 & similar & 5  & 7 \\
  download & get file &  & to 2021 &  &  &  &  \\
  %download &  &  & 2021 &  &  &  &  \\
 \midrule
  logistics [fedex] & shipping cost paid & 5 & 2018 & 18.0 & similar & 1 & 5 \\
  click & view status order &  &  &  &  &  &  \\
  %click &  &  &  &  &  &  &  \\
 \bottomrule
\end{tabular}  
%}
\end{center}
\end{table*}

\subsubsection{Multi-layer Clustering in Four Steps}
We followed a systematic approach that consisted of four steps to obtain clusters of similar emails  and was done automatically using \texttt{Python Dataframes}, making it easy to replicate.
\textit{Step~1:} All emails are initially grouped into a single cluster, then segregated into sub-clusters based on the `From-Sector'. This is because emails within a campaign, despite variations, share a common origin sector. For instance,  two emails from different sectors like `financial' and `email' would not belong to the same campaign. \textit{Step~2:} The clusters are further divided based on the `Action-Generic' code, as emails part of a single attack usually have the same attack vector. \textit{Step~3:} The obtained clusters are divided based on the impersonated organization's name. Although in the previous section, we discussed possible campaigns with different organization names, it is still a discriminating factor in identifying similar emails. This resulted in a total of 104 clusters. Out of these, 36 contained more than one email and 14 clusters had more than 5 emails. If we exclude the singleton clusters, the mean cluster size is 11.25, and the median is 4.00. It is important to note that singleton clusters are very common in campaign research. \textit{Step~4:} We divide the clusters based on the `Main Topic' and `Action-Specific' codes. These codes summarize the main `what' and `why' of the phishing email, and hence are very important to identify campaigns. For the sake of this illustration, we used an exact phrase match to divide the clusters. Since these are open-ended codes with many possible variations, this matching criteria is very specific, and more advanced alternatives like topic models, Levenshtein distance (or edit distance), or contextual similarity could be used.

\subsubsection{Analysis of clusters}
We pick five clusters from the resultant clusters and manually analyze them in detail. The summary of these results is presented in Table~\ref{tab:qual_camps}. These findings illustrate how incorporating qualitative attributes enables us to identify patterns and similarities among phishing emails within a given campaign, even when technical features such as URLs or header features vary. Remarkably, three of the five clusters identified occurred over multiple years, yet existing detection algorithms failed to filter them out despite being in the same language and sharing similar wording and layout. Visually, all emails within a cluster were nearly identical with minor differences. Furthermore, we calculated the average pairwise Levenshtein Distance~\cite{levenshtein1966} between all subject lines, as this method is commonly used to quantify similarities between text. The distance between two phrases is the minimum number of single-character edits required to change one phrase into the other. The findings indicate that only one cluster contained emails with identical subject lines, and the rest had high variation. Another crucial observation was the existence of multiple usernames and domains in the From headers and URLs within a cluster. Although URL-based features are frequently used for phishing campaign detection, emails within the same campaign can exhibit variations in these aspects. It is important to highlight that our proposed codes not only enhance campaign identification but also have the potential to improve the explainability of the algorithms and resultant clusters. By providing a clear set of characteristics that define each cluster, as shown in Table~\ref{tab:qual_camps}, it becomes easier to explain and interpret the results. We will not discuss in detail one cluster, emails claiming to be from the United Services Automobile Association~(USAA), an American financial service.

\begin{figure*}[h!]
\begin{subfigure}{0.5\textwidth}
\includegraphics[width=\textwidth]{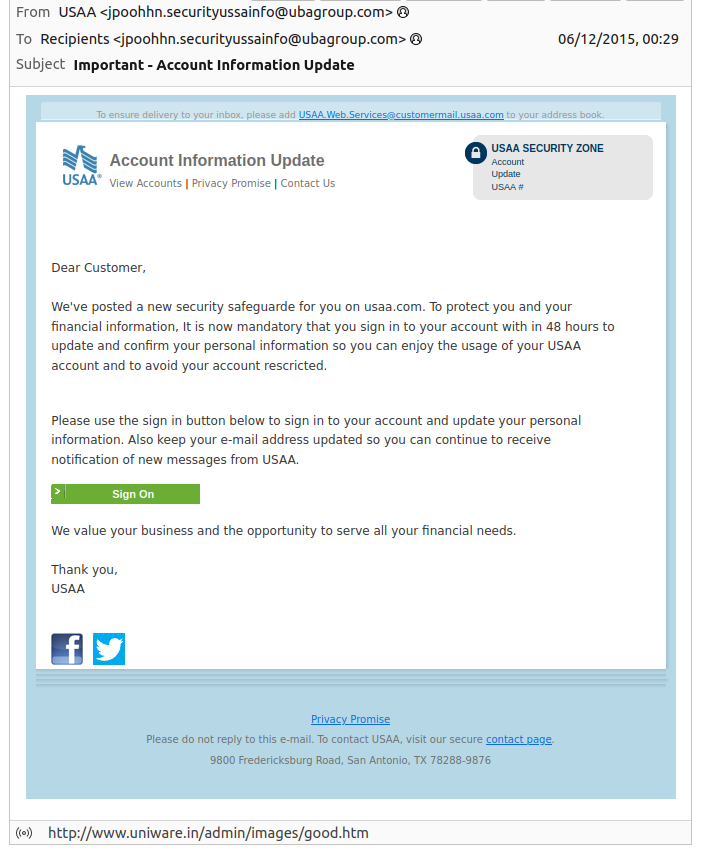} 
\caption{USAA Sample 1}
\label{fig:usaa_em_1}
\end{subfigure}
\begin{subfigure}{0.5\textwidth}
\includegraphics[width=\textwidth]{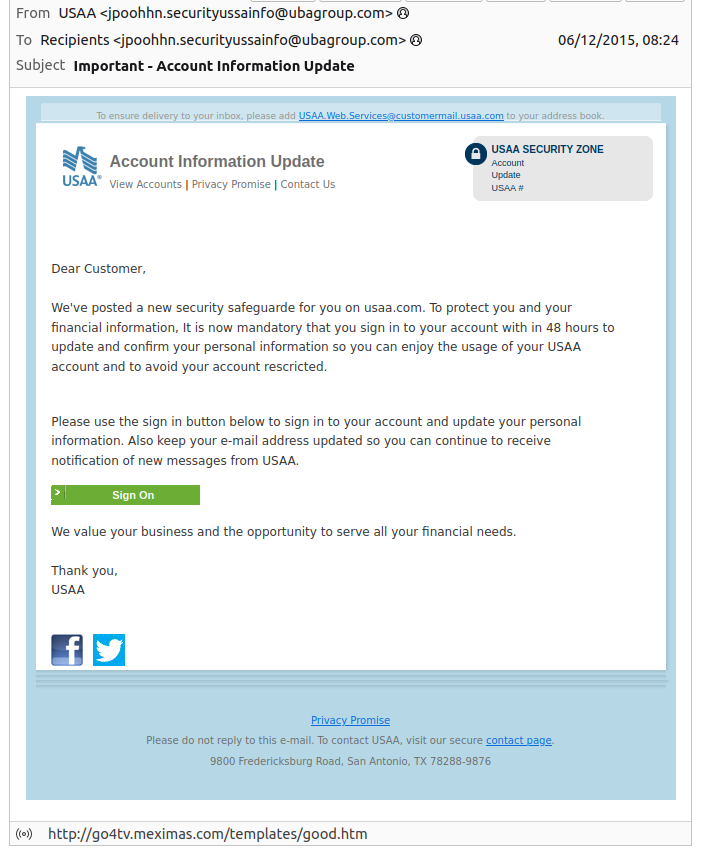}
\caption{USAA Sample 2}
\label{fig:usaa_em_2}
\end{subfigure}
\begin{subfigure}{0.5\textwidth}
\includegraphics[width=\textwidth]{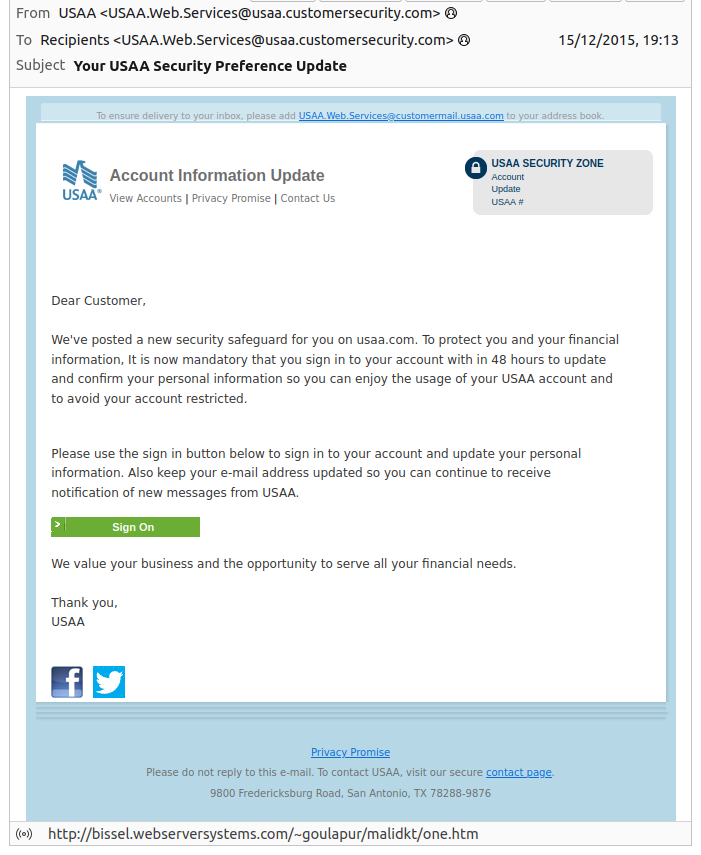}
\caption{USAA Sample 3}
\label{fig:usaa_em_3}
\end{subfigure}
\begin{subfigure}{0.5\textwidth}
\includegraphics[width=\textwidth]{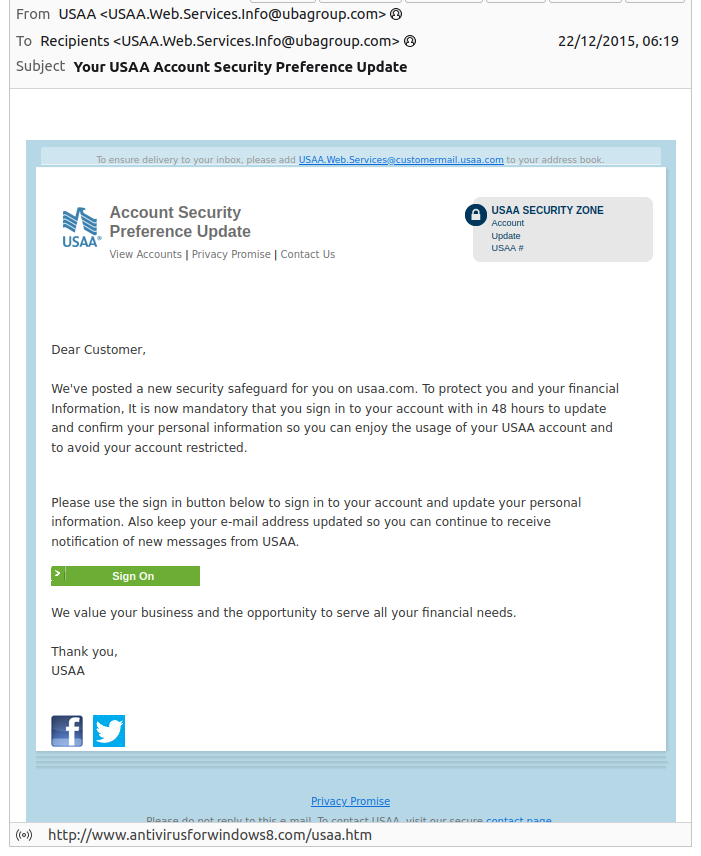}
\caption{USAA Sample 4}
\label{fig:usaa_em_4}
\end{subfigure}
\caption{Emails from the USAA cluster.}
\label{fig:usaa_campaign}
\end{figure*}

This cluster of phishing emails claim to be from the United Services Automobile Association~(USAA), an American financial services company providing insurance and banking products. The emails claims that USAA has implemented a `new security safeguard' and ask the user to sign-on to `update personal information'. As shown in Figure~\ref{fig:usaa_campaign}, all four emails are an exact visual and textual match. All four identified emails are from a single year.
Two of the emails (Figures~\ref{fig:usaa_em_1} and \ref{fig:usaa_em_2}) were sent on the same day and share the same sender name and email address but they have different URLs. At the time of this study, the URLs were inactive and hence further analysis was not possible. The other two emails (Figures~\ref{fig:usaa_em_3} and \ref{fig:usaa_em_4}) were sent in the same month but had different sender addresses and URLs. In Table~\ref{tab:usaa}, we provide a comparative analysis of some of the commonly used machine-readable features. Note that the first IP address are the same for emails~\ref{fig:usaa_em_1} and~\ref{fig:usaa_em_2} and emails~\ref{fig:usaa_em_3} and~\ref{fig:usaa_em_4}. The first IP address was extracted from the last `Received' header. Received headers show the path that a message takes from the message sender to the final recipient. The last one shown should be the one where it left the sender, and hence in considered a reliable feature. So the above pairs should have been sent from the same server even though their sender address and URL domains are different. Without ground truth, we can't be certain if all the emails from this cluster originate from the same source, however for many use-cases, such as identifying authorship or providing response to reported phish, this cluster would be valuable. 

\begin{table*}[h!]
    \centering
    \begin{tabular}{|c|c|c|c|c|c|c|c|c|}
        \hline
        ID & Sender Name & Sender Address and Return Path & First IP Address & Main URL - Domain & DKIM \\
        \hline
        \ref{fig:usaa_em_1} & USAA & jpoohhn.securityussainfo@ubagroup.com & 173.221.126.99 & uniware.in & none \\
        \ref{fig:usaa_em_2} & USAA & jpoohhn.securityussainfo@ubagroup.com & 173.221.126.99 & go4tv.meximas.com & none \\
        \ref{fig:usaa_em_3} & USAA & usaa.web.services@usaa.customersecurity.com & 70.43.42.42 & bissel.webserversystems.com & none \\
        \ref{fig:usaa_em_4} & USAA & usaa.web.services.info@ubagroup.com & 70.43.42.42 & antivirusforwindows8.com & none \\
        \hline
    \end{tabular}
    \caption{Analysis and comparison of USAA clusters}
    \label{tab:usaa}
\end{table*}

\subsection{Automatic Response to Users}
\label{sec:app2}
In Section 2.2, we discuss the importance of automatic response or guidance tools to support end-users in detecting phishing emails. Prior research on in-the-moment phishing guidance has predominantly focused on the use of security warnings, commonly in the form of browser-based or banner warnings~\cite{egelman2008you}. \textit{Browser warnings} typically appear after a user has already clicked on a link, which may sometimes be too late. On the other hand, \textit{banner warnings} alert users to the suspicious nature of an email as soon as it is opened. However, these warnings often fail to explain specific suspicious elements, placing the responsibility of identifying potential threats on the user. A recent study by Petelka\textit{ et al.}~\cite{petelka2019put} revealed that warnings focused on links reduce phishing click-through rates compared to email banner warnings, with forced attention warnings being the most effective. However, as organizations receive emails from various domains and phishers increasingly employ URL obfuscation techniques such as redirection and shortening, relying solely on URLs is insufficient.
Consequently, an overall email summary that contextualizes the content could greatly enhance decision-making for individuals. Recent studies suggest that future user guidance should leverage email context to improve efficacy~\cite{jenkins2022phished, kashapov2022email, papathanasiou2023cognitive}. To achieve this, we need a robust method for modeling email context, which the phishing codebook provides. Through this qualitative study, we identified a set of descriptive features from phishing emails that can be used to create well-tailored responses by explaining the scams and tricks being used in the email and drawing the user's attention to crucial indicators.

The `From-Sector' code helps categorize the email into a type of scam and craft a better explanation using this classification. For instance, the scam explanation for a `Document Sharing' scam and a `Financial/Banking' scam are very different. Similarly, the `Action-Generic' code is crucial in creating correct advice because the appropriate response to a phishing email can vary significantly depending on the action requested. For instance, a phishing email urging recipients to click on a malicious link may necessitate different remedial measures compared to one that encourages users to download a malicious attachment. The `Company Name' impersonated in the email allows for cross-referencing with the URL and sender domain. Exposing such inconsistencies can help provide users with confidence in the guidance. `Threatening Language' and `Urgency Cues' within the email content serve as a flag for potential manipulation, guiding users to be cautious and verify the legitimacy of requests before taking any action. Such contextual advice can not only improve the users' decision-making process but can also educate them for future attacks. 

\section{Discussion}
\label{sec:disscussion}
In the previous section, we have demonstrated the practical application of the proposed Phishing Codebook for two crucial phishing mitigation tasks. We will now discuss other potential applications of the codebook.

%Firstly, we illustrated how the codebook's output can be used to identify similar phishing emails, likely from the same campaign. This can significantly enhance the IT staff's ability to manage reported phishing incidents efficiently. Secondly, we discussed how the coded representation of the email can be used to create well-tailored responses to end-users and provide them with guidance on how to respond appropriately. Our experiments showed that using qualitative features to identify phishing campaigns is an effective approach that can identify patterns and similarities among phishing emails within a given campaign, even when technical features such as URLs or header features vary. 

\subsection{Representative Feature Set}
The proposed codebook contains eight high-level codes that capture the human-focused context of phishing emails. The codes ``From - Company Name" and ``From - Sector" offer insights into the impersonated organization, while ``Salutation" indicates the presence of personalization in the emails. ``Threatening Language" and ``Urgency Cues" highlight linguistic cues commonly used to influence the recipient's reaction and form an essential part of the scam. ``Action - Generic" and ``Action - Specific" describe the type of action being solicited from the recipient and the reason for the same. Finally, ``Main Topic" summarizes the underlying scam, providing a concise description. The feature set in our codebook demonstrates a balanced approach to representativeness, incorporating various aspects of phishing emails while ensuring comprehensiveness by covering key dimensions relevant to the underlying scam of the phishing email. In the exploratory phase, we assessed other features like images, logos, and disclaimers. However, we encountered challenges in defining sub-codes for these elements, as categorizations such as `good', `mediocre', or `bad' proved to be highly subjective and posed difficulties in maintaining consistency among coders. We prioritized consistency and coding reliability for our feature set choice.

%\subsection{Identifying phishing campaigns}
%Email text-based features are commonly considered important in identifying email campaigns or clusters since attackers often send bulk emails with similar text and wording, using templates or kits. In Section~\ref{sec:bg}, we discuss in detail various use-cases for grouping malicious emails. In this paper, we attempted to define and extract a set of human-facing features of phishing emails that could be used for campaign detection. The features described in our Phishing Codebook are: `From-Company Name', `From-Sector', `Salutation', `Threatening Language', `Urgency Cues', `Action-Generic', `Action-Specific', and `Main Topic'. These descriptive features mimic the thought process of an average user when processing suspicious emails. Despite the common belief that campaign emails share the same underlying scams and structures, there has been limited prior research on finding efficient ways to represent and utilize these features. Our codebook represents the first attempt to address this gap in the literature. To demonstrate the importance of such qualitative features, we performed a multi-criteria classification task to identify some notable phishing clusters, as shown in Section~\ref{sec:app1}. Our illustration clearly emphasizes the potential of incorporating qualitative features in detecting meaningful phishing and spam groups.

\subsection{Structured email representation}
As part of the coding process, the authors manually analyzed 503 phishing emails. We observed that the content of an email can vary greatly depending on its purpose and context, ranging from short and simple messages to lengthy and detailed reports. While some emails include elements such as salutations, signatures, and disclaimers, others completely lack context or information. For instance, the shortest email in our coded dataset contained a mere 27 characters, whereas the longest email had a staggering 15,488 characters. Saka, Vaniea, and Kokciyan~\cite{saka2022} speculate that an overlap in wording, caused by generic text could cause poor scam clustering when using language models such as BERT embeddings to represent unstructured text. Han and Shen~\cite{han2016} also emphasize that when two spear phishing emails from different campaigns share similar features, the detection algorithm could be overwhelmed. Furthermore, they specify that when emails are extremely short, such as one sentence or several phrases, text features become much less stable and less informative to classify emails as phish or legitimate. Such variation and discrepancy in email structure can pose a challenge for various Machine Learning and NLP algorithms, which are widely used for campaign or cluster identification. To address this variability, our Phishing Codebook offers a standardized framework for extracting key information from emails and presenting it in a clear and concise manner. By normalizing highly variable emails into a standard representation, we can improve the effectiveness of phishing mitigation techniques. With advancements in semi-supervised learning~\cite{van2020survey} and language models~\cite{bowman2023}, such structured information extraction can be achieved.

\subsection{Creating labelled datasets}
Organizational phishing mitigation techniques often require the classification of potential phishing threats into clusters. For example, some phishing emails make it past the filters, get reported by users, and then must be processed by IT staff. This can be a challenging task, particularly if there are numerous such emails to deal with~\cite{saka2022,althobaiti2021}. Clustering can also be done to understand trends over time and to gather evidence about common authors~\cite{seifollahi2017, alazab2013}. Detecting botnets or common sender sources is also important and can require emails to be clustered automatically~\cite{wei2008, han2016}. Such email grouping is a challenging task owing to its inherently unsupervised nature. The evaluation of phishing email clusters is currently done either through automatically detectable features, such as URLs or using researcher-generated labels, which often lack proper documentation. The absence of labels necessitates researchers to adapt by either curating a label set or manually verifying the outcomes to evaluate the performance of their model. Currently, there is no standard or reliable way to do so. For instance, two recent studies~\cite{saka2022, sheikhalishahi2020} employed a manual approach to analyze phishing emails and assign labels based on different textual attributes. The absence of a standardized method for this process presents a significant challenge in terms of reproducibility and comparability. Inconsistencies in labeling data can make it challenging to compare multiple models and determine the best one. The Phishing Codebook provides a framework for researchers to use as a reference to systematically code phishing emails and assign qualitative labels. By using a uniform approach to label data, researchers can improve the consistency of results and reduce the ambiguity associated with comparing outcomes across different studies.

\subsection{Recommendations for Security Workers}
In Section 5, we identify key challenges in phishing emails, both individually and in clusters. One significant issue is the misuse of legitimate services to distribute phishing content, making emails appear trustworthy while containing malicious elements. Information security officers must proactively educate employees, especially those with less technical knowledge, through continuous real-time training rather than traditional methods. Long-lasting scams highlight the limitations of current technical filters that rely on blacklists and URLs, necessitating the integration of qualitative attributes into detection strategies. Variations in phishing campaigns, such as minor letter changes and different links, demand enhanced rule-based filters incorporating organization-specific names and topics. Ensuring organizational security requires a multi-layered approach, combining robust technical filters and servers with effective training programs. This empowers end users to identify and mitigate threats, reinforcing their role in maintaining email security daily.

\subsection{Limitations}
Although the authors built the codebook based on a thorough discussion and deliberation, it is possible that certain features were overlooked or sub-optimally represented. There are additional features from other parts of the email, such as attachments, URLs, and images, that can also be useful, which we intend to explore in the future. While the features in our codebook may not be exhaustive, we hope it serves as a good starting point for future researchers. Additionally, the publicly available dataset used in this research represents a single user's personal inbox and thus does not reflect the frequency and scale of attacks an organization usually receives. However, the authors applied the codebook to an independent phishing email dataset to ensure its generalizability (Section~\ref{subsec:independent}). Despite these limitations, we hope this study will serve as a foundation for future researchers.

\section{Conclusion and Future Work}
\label{sec:conc}

In this paper, we conducted a qualitative analysis of phishing emails to identify their qualitative traits and their role in identifying phishing campaigns and clusters. 
The main motivation of this study was to create a better understanding of phishing emails by dissecting their complex structure and generating a human-centric representation of emails using features that users naturally use to judge suspicious emails. We conducted iterative qualitative coding of phishing emails and created a Phishing Codebook consisting of eight high-level codes, which serve as a reliable framework for extracting relevant information from phishing emails~(RQ1). Using this framework, we coded a set of 503 phishing emails with high inter-rater reliability. The study further provides interesting insights into the various challenges posed by phishing emails independently and as a group~(RQ3). Furthermore, we illustrate the utility of the Phishing Codebook to cluster emails through a simple multi-criteria classification experiment~(RQ2). In our analysis, we found that even similar phishing emails exhibited variations in subject lines, usernames, domains, and URLs. This experiment highlights the potential of incorporating descriptive and qualitative features for campaign detection. As a second use case, we also discussed how a standard representation of emails can be useful in creating well-tailored responses to assist end-users in responding to phishing appropriately. To the best of our knowledge, this is the first attempt to create such a framework that aims to model the context of phishing emails by categorizing their qualitative traits and defining a novel set of descriptive features.

The results of this study provide various directions for future research. Our immediate future goal is to automate the coding process and create an intelligent tool to automatically extract important information from phishing emails by considering human-centric features as defined in our codebook.
In addition to automating the coding process, future work involves augmenting the codebook with elements about user pretext and tailoring to subject/organization. These aspects can add immense value to the Phishing Codebook. 
Furthermore, we plan to conduct various experiments on downstream applications of this codebook. This includes the identification of phishing campaigns (an illustration is provided in Section~\ref{sec:app1}), detection of phishing emails, and creation of auto-response tools that provide instantaneous analysis and guidance of emails to humans: (i) end-users who may need guidance to decide on how to interact with a particular email, and (ii) IT support teams to implement mitigation strategies to protect users and organizations.

\bibliography{references}

\section*{\textbf{Appendix A: Phishing Codebook}}
\label{app:app_codebook}
This codebook is a framework to extract important information from phishing emails in a structured format. While coding only consider the user-facing part of the email; subject line, displayed header, and the main body visible to end-users. If there are multiple codes for a single aspect, commas should be used to separate them.

\subsection{From – Company Name}
If there is a named company in the from address, subject line, or email body that the email claims to be from. State the name. For example, “PayPal”. If two companies are named, include both. Only code for companies or organizations, ignore the reference to a named person. If there are two variants of the same company, name both with the more recognizable name first and separated by commas. For example, ‘Microsoft, Outlook’. If there is no name specified in the email, then use one of the following:  

\begin{itemize}
    \item Monkey – The email explicitly says it is from some group associated with monkey.org.
    \item Organization – The email is claiming or implying to be from an organization the user works for. They may use generic terms like “HR” or "Manager” that are associated with a company. Often the organization is not named, but word usage implies that the email is internal.
    \item None - The email makes no claim about who it is from, or the coder cannot infer a specific organization name from it. The email address may just look like garbage.
\end{itemize}

\subsection{From – Sector}
The type of sector that the email claims to be from. Refer to the list below to code this. If a company is involved in several different types of sectors, code for the one that the email content refers to. For example, Amazon is both a Shopping website and a Logistics service, if the email is about package delivery, then Logistics should be coded and not Shopping.

\begin{itemize}
    \item Financial – banks, credit cards, investment company, cryptocurrency
    \item Email – email provider, Outlook, or a department that manages email accounts such as the IT team within a larger organization that may not be email focused.
    \item Document share – Online service that allows users to share documents with each other. For example: DocuSign or OneDrive.
    \item Logistics – shipping and delivery of goods and parcels
    \item Shopping – regards to purchasing goods or services online, companies like Amazon.
    \item Service provider – An organization that provides online services and does not fall into any sector described above. So, a generic email claiming to be from Apple should be coded here, but an issue with the Apple hosted email should be coded under "Email".
    \item Security – any company that claims to be a security provider, anti-virus service, or regarding identity protection due to a security breach.
    \item Government – any email from a government organization. For example, tax-related emails, HMRC, or VISA-related emails, and so on.
    \item Unknown – the coder can’t tell from just the email or does not know what the company is.
\end{itemize}

\subsection{Salutation}
The type of salutation used to address the user. Some phishers tailor emails to users by using their name or email to make it seem legitimate, while others send out mass emails with no specific salutation. Code if the email addresses the user specifically, either in a proper salutation or in the email body text. IGNORE the email header for this code.

\begin{itemize}
    \item Name - if the name of the user is used. For example: ‘Dear Jose’, ‘Attention Jose’
    \item Email - if user email id is used in the salutation
    \item Generic - if no name or personal salutation is used. For example: ‘Dear User’
    \item None - if there is no salutation at all or a reference to an individual. For example: ‘Hello’
\end{itemize}

\subsection{Threatening language}
Phishers often use a threat or warning to scare users into taking action. Code if the email subject line or message contains any threatening language or tone. This includes talking about negative consequences or loss.

\begin{itemize}
    \item Threat - If there is a DIRECT threat statement or wording in the email. For example: “account will be deleted” or “money will be lost”.
    \item None- If the email has no threat or incentive mentioned. Or if you cannot accurately infer one from the email. Also includes cases where the outcome of a process is described (I.e. mail being held) which may be undesirable, but no direct threat statement exists.
\end{itemize}

\subsection{Urgency Cues}
Phishers often try to create a sense of urgency to encourage, or even demand, immediate action in a bid to fluster the receiver. Code if the email message or subject line contains time pressure or urgency cues, including implied.

\begin{itemize}
    \item Urgent – if there is a mention of time limit or mention any urgency words. For example: ‘files will be lost in 24hrs’ or ‘expire in 3 days’ or ‘click Immediately’ or ‘soon’.
    \item None – If there is no such time pressure scare or wording used in the email. Or if you cannot accurately infer from the message.
\end{itemize}

\subsection{Action – Generic}
Code the action being prompted in the email. This only includes clearly stated explicit actions, and not any implied actions.

\begin{itemize}
    \item Click – if the email is asking you to click on any links
    \item Download – if the email is asking you to download an attachment or application
    \item Reply/Email – if the email is asking you to reply or send an email to a given address
    \item Call – if the email is asking you to call a number
    \item Other – If any other action is mentioned that is not covered above.
    \item None – No clear action is requested, or the coder is unsure what is being asked for.
\end{itemize}

\subsection{Action – Specific}
Provide details regarding the reason the phisher has given for the action. We will use in vivo coding for this column - copy and paste a few words/phrases directly from the email text. For example: For the action ‘click link to verify account’, the code is ‘to verify account’.

\subsection{Main Topic}
Code the main purpose of the email. Use a few words or a phrase from the email to summarize the main topic. For example: ‘package has been returned’.

%------------------------------------------------------------------------------------------------------

\end{document}